\documentclass[12pt]{article}
\begin{document}
\newcommand{\bea}{\begin{eqnarray}}
\newcommand{\eea}{\end{eqnarray}}
\newcommand{\ba}{\begin{array}}
\newcommand{\ea}{\end{array}}
\newcommand{\spsigma}{\begin{array}[t]{c} \mbox{Sp} \vspace{-1ex} \cr
\mbox{$\scriptstyle{(\sigma)}$} \end{array} }
%
%
\phantom{A}
\vskip 5mm
\begin{center}
\section*{\Large\bf
Fermions and Disorder in Ising and Related Models in  
Two Dimensions
}
\vskip 3mm
{\bf V.N.~Plechko }\\
\vskip 3mm
{Bogoliubov Laboratory of Theoretical Physics,}\\
{Joint Institute for Nuclear Research, 141980 Dubna, Russia}
\end{center}
\vskip 5mm
\begin{abstract}
The aspects of phase transitions in the two-dimensional Ising models
modified by quenched and annealed site disorder are discussed in the
framework of fermionic approach based on the reformulation of the problem
in terms of integrals with anticommuting Grassmann variables. 
\end{abstract}
\vskip 8mm
{\renewcommand{\thefootnote}{*}
\footnotetext{ \ 
The present discussion is partly based on a talk given at the International
Bogoliubov Conference on Problems of Theoretical and Mathematical Physics,
MIRAS--JINR, Moscow--Dubna, Russia, August 21--27, 2009.}}
\setcounter{footnote}{0}

%
%
%
\section{Introduction}

The two-dimensional (2D) Ising model (2DIM) plays important role in the
theory of phase transitions and critical phenomena  due to the analytic
results available (in pure case) over the whole temperature range
\cite{ons71,sml64,mpw63,hugr60}.  In this report, we review the new
mathematical methods of analysis and the results so far obtained for 2DIM
modified by site disorder with application of the anticommuting (Grassmann)
integrals.  The Ising model by itself, in its original formulation, is a
lattice model of a ferromagnet presented by a set of Ising spins
$\sigma_{mn}=\pm1$ interacting with their nearest neighbours along the
lattice bonds \cite{ons71,sml64,mpw63,hugr60}.  The modern approaches to
Ising models are merely based however on the fermionic path integral
reformulation of the problem in terms of the integrals with anticommuting
Grassmann variables \cite{ber69,sam80,itz82,ple85d,ple85t, ber66}. The
advantage of the use of Grassmann variables is that they are canonical
variables, as distinct from Ising spins, so that one can pass to the
momentum space for fermions \cite{ber69, sam80, ple85t, ber66}.  In
the  pure case, the fermionization of 2DIM results the Gaussian fermionic
integral for the partition function, $Z$, which in essence means the exact
solution of the problem \cite{ber69,sam80,ple85d,ple85t}. The
formulations of this kind also admit the interpretation of the 2D Ising
model as a lattice quantum field theoretical (QFT) problem \cite{dots83,
shal94, jugshal96, ple98}. In particular, the pure 2DIM on a rectangular
lattice may be presented by the Majorana action with two-component massive
fermions on a lattice \cite{ple98}. By doubling the number of fermions, one
can pass as well to the Dirac action \cite{ple98}.  The effects of disorder
in 2DIM have been extensively studied during last decades both
theoretically and in the precise Monte-Carlo simulations
\cite{dots83}-\cite{lawri84}.  The disordered versions of 2DIM may assume
either random modification of the interaction along the lattice bonds
\cite{dots83,shal94, raja98, kim00, ziegler90, nihat09, wuzhao09}, or an
admixture of the random nonmagnetic impurities at lattice sites
\cite{ple98, shvas01, ssvas98, martpla07, hasen08, kenaruiz08, cckenrul09},
also see \cite{beg71, beale86, deng05, silva06, cfple08}. In each of
this cases one may be interested, motivated by physical considerations, and
possible applications, in either quenched or annealed versions of disorder.
In the quenched version, the impurities are assumed to be frozen over the
sample.  In this case their distribution does not depend on temperature and
other tuning parameters, like magnetic field, and one have to average
rather the free energy $-\beta F =\ln Z$ than the partition function $Z$
itself, over the impurities \cite{dots83, shal94, jugshal96, ple98}.  In
the annealed version, the impurities may be created and annihilated by a
variation of external parameters and their concentration is governed by the
temperature rate and associated chemical potentials \cite{beg71, beale86,
cfple08, lawri84}.  In this case, one has to average in $Z$ itself over all
states \cite{beale86, deng05, silva06, cfple08}.  In essence, for annealed
site disorder, the dilute site can be viewed as being presented by
additional (zero) component of Ising spin. The resulting model is also
known as the spin-$1$ Ising model, or the Blume-Capel model \cite{beg71,
beale86, cfple08, lawri84}.  The basic variable in the Blume-Capel model is
$S_{mn}=0, \pm1$. The modifications introduced by disorder of any kind
typically result in the appearance of the additional non-Gaussian terms in
fermionic action.  Despite of the non-Gaussian action in the fermionic
integral for $Z$, the precise results can still be derived for disordered
2D Ising models \cite{dots83,shal94, jugshal96, ple98,cfple08}. In what
follows, we only consider the generic case of the random-site disorder
(site dilution) introduced by adding some amount of nonmagnetic impurities
into a sample, which may be either quenched or annealed, and discuss the
consequences that can be derived from the fermionic integral
representations for the partition functions of that models.

%
%
%
\section{The quenched site dilute Ising model}

The basic variable in the pure 2DIM is the dichotomic Ising spin
$\sigma_{mn} =\pm1$. The spins are  disposed at the sites of a regular
two-dimensional lattice and interact with nearest neighbours along the
lattice bonds. The disordered version (quenched site dilution) assumes
that some sites may be nonmagnetic at random. It is suitable to introduce
such sites by adding the variable $y_{mn}=0,1$ at each $mn$ site,
corresponding to the magnetic moment of Ising spin at a given site
\cite{ple98}. The resulting hamiltonian is:

\bea
H\{y\,|\,\sigma\} = -\sum\limits_{mn}^{}
\left[\,J_{1}\,y_{mn}y_{m+1n}\sigma_{mn}\sigma_{m+1n}
+J_{2}\,y_{mn}y_{mn+1}\sigma_{mn}\sigma_{mn+1}\right]\,,
\label{ham1ah}
\eea
where $J_{1,2}$ are the ferromagnetic exchange energies; the lattice sites
are marked by discrete coordinates $mn$, where $m,n=1,2,\ldots,L$, are
running in horizontal and vertical directions, respectively; we put
$L^2\to\infty$ at final stages.  For fixed disorder, the partition function
and free energy are defined by the canonical equations: $Z\{y\}\, =\Sigma
\exp(-\beta H\,\{y\,|\, \sigma\}) =\exp(-\beta F\,\{y\})$, where the sum is
taken over the all possible spin configurations provided by $\sigma _{mn}
=\pm1$ at each site. The hamiltonian modulus in the Gibbs
exponential is:

\bea
-\beta H\,\{y\,|\,\sigma\} =\sum\limits_{mn}^{}
\left[\,b_{1}\,y_{mn}y_{m+1n}\sigma_{mn}\sigma_{m+1n}
+ b_{2}\,y_{mn}y_{mn+1}\sigma_{mn}\sigma_{mn+1}\right]\,,
\label{ham1yy}
\eea
where $b_{1,2} =\beta J_{1,2}$,  and $\beta=1/kT$ is the inverse
temperature. For a typical bond weight from $Z$ we write:
$\exp\,(b\,yy' \sigma\sigma') =\cosh(b\,yy') +yy' \sigma \sigma' \sinh(b)$,
since $\sigma\sigma' =\pm1$ and $yy'=0,1$. The partition function can then
be written in the form:  $Z\,\{\,y\,\}=R\,\{\,y\,\} \,Q\, \{\,y\,\}$, where
$R\{y\}$ is a nonsingular spin-independent prefactor, formed by a product
of $\cosh(byy')$, while $Q\{y\}$ is the reduced partition function:

\bea
Q\{y\}=\spsigma\!
\Big\{\prod\limits_{\,mn}\;
(1+t_{1}\,y_{mn}y_{m+1n}\sigma_{mn}\sigma_{m+1n})
(1+t_{2}\,y_{mn}y_{mn+1}\sigma_{mn}\sigma_{mn+1})\Big\}\,,\;\;
\label{qss1}
\eea
where $t_{1,2} =\tanh\,b_{1,2}$, and we assume a properly normalized
spin averaging, such that $\mbox{Sp}\,(1)=1$ and
$\mbox{Sp}\,(\sigma_{mn})=0$ at each site.
In given case, since we are interesting in
quenched disorder, we have to average over $y_{mn}=0,1$ rather the free
energy $-\beta \ln Z\{y\}$ than the partition function $Z\{y\}$ itself.
The prefactor $R\{y\}$ provides only additive nonsingular contribution like
$\ln R\{y\}$ to $\ln Z\{y\}$ and will be ignored in what follows. The
problem thus reduces to the averaging of $-\beta F_Q =\ln Q\{y\}$ over
$y_{mn}=0,1$ at each site.
\footnote{ \
The situation is different for the annealed case (the Blume-Capel model),
where the zero state provided by $y_{mn}=0,1$ is rather to be considered
as a zero component of the BC spin $S_{mn}=0,\pm1$ so that one have to
average over the all three states of $S_{mn}=0,\pm1$  directly in $Z$,
inside of the logarithm. Respectively, the  cosine factors (product
$R\{y\}$) are now to be preserved under the averaging in $Z$. These factors
in fact add new degrees of freedom for clusterization of the magnetic sites
at low temperatures, and are eventually responsible for the appearance of
the tricritical point in the Blume-Capel model at strong dilution
\cite{cfple08}. In fermionic language, the tricritical point is
associated with vanishing of the kinetic (stiffness) coefficient
in the Blume-Capel fermionic action at strong
dilution \cite{cfple08}.}
The known device to avoid the averaging of the logarithm is the replica
trick: $[-\beta F_{\,Q\,}\{\,y\,\}] =[\ln\,Q\{\,y\,\}] =[\frac{1}{N}\,
(Q^N\{y\}-1) ]_{N\to0}$, where $[\ldots]$ stands for the average over the
impurities. In this scheme, one takes $N$ identical copies of the original
partition function and average $Q^{\,N}\{y\}\,$, with formal limit $N\to0$
to be performed at final stages. The simplest distribution for the
averaging over the impurities is assumed in what follows: $w\,(y_{mn}) =
p\,\delta\,(1 - y_{mn}) + (1 -p)\,\delta\,(y_{mn})$, where  $p$ is the
probability that any given site, chosen at random, is occupied by the
normal Ising spin, while $1-p$ is the probability that the given site is
dilute. The averaging of any function like $A(y_{mn})$ then results:
$[A(y_{mn})]=p\,A(1) +(1-p)\,A(0)$.

The partition function with fixed disorder (\ref{qss1}) can be transformed
into a Gaussian fermionic integral following the method of the
mirror-ordered factorization for the density matrix
\cite{ple85d,ple85t,ple98}.
Introducing a pair of fermionic (Grassmann) variables $a_{mn},a_{mn}^{*}$,
we write for the horizontal weight:
\bea
&&
1+t_{1}y_{mn}^{}y_{m+1n}^{}\sigma_{mn}\sigma_{m+1n}
\cr
&& =\int\limits_{}^{}da_{mn}^{*}da_{mn}\,\mbox{e}^{\,a_{mn}a_{mn}^{*}}\,
(1+a_{mn}^{}y_{mn}^{}\sigma_{mn}^{})\,
(1+t_{1}\,a_{mn}^{*}y_{m+1n}^{}\sigma_{m+1n}^{})\,. \;\;\;
\label{fact1}
\eea
In a conventional notation, the horizontal weight is now presented as
a product of two factors, $A_{mn}A_{m+1n}^{*}$, with decoupled spins,
taken under the Gaussian averaging.
 \footnote{ \
Let us remember that Grassmann variables (nonquantum fermionic fields) are
the purely anticommuting fermionic symbols. Given a set of Grassmann
variables $a_1,a_2,...\,, a_N$, we have $a_ia_j+ a_ja_i =0$, and $a_{j}^{2}
=0$.  The rules of integration over Grassmann variables (fermionic path
integral) were originally introduced by F.A. Berezin in QFT context
\cite{ber66}. The elementary rules of integration for one variable are
\cite{ber66}:  $\int da_j \cdot a_j = 1\,,\,\int da_j \cdot 1=0$. In a
multidimensional integral, the differential symbols $da_1,da_2,
\ldots,da_N$ are again anticommuting with each other and with the
variables. Gaussian fermionic integrals are in general related to the
determinants and Pfaffians. In the field-theoretical language, the
fermionic form in the exponential under the integral is typically called
action. For more comments about Gaussian fermionic integrals in a related
context also see \cite{ber69, sam80, ple85t, ber66, ple98, cfple08}.}
In a similar way,
one prepares the factorized vertical weights in the
form $B_{mn}B_{mn+1}^{*}$. At next stage, we have to arrange the factors
in their global products in order the elimination of spin variables be
possible \cite{ple85d,ple85t, ple98}. The final point is that we have to
average over $\sigma_{mn}=\pm1$ the product of four factors with the same
spin like $A_{mn}^{*}B_{mn}^{*}A_{mn} B_{mn}$ at each site, thus passing to
a purely fermionic expression. This results the integral \cite{ple98}:
\bea
&& Q\{y\} = \int
\prod\limits_{mn}^{}db_{mn}^{\,*}db_{mn}^{}
da_{mn}^{\,*}da_{mn}\exp\sum\limits_{mn}^{}
\Big\{ a_{mn}^{}a_{mn}^{\,*}+b_{mn}^{}b_{mn}^{\,*}\,+
\nonumber\\
&& +\, y_{mn}^{2}\Big[\,a_{mn}^{}b_{mn}^{}+ \,t_{1}t_{2}\,
a_{m-1n}^{\,*}b_{mn-1}^{\,*}+\,(t_{1}a_{m-1n}^{\,*}
+\, t_{2}b_{mn-1}^{\,*})(a_{mn}^{}+b_{mn}^{})\,\Big]\Big\}\,,\;\;\;
\label{qab1}
\eea
where $a_{mn}, a_{mn}^{\,*}, b_{mn}, b_{mn}^{\,*}$ are Grassmann variables.
The disorder parameters $y_{mn}^{2} =0,1$ are still free parameters in the
above integral, while Ising degrees being already eliminated. In turn,
integrating out a part of fermionic variables from (\ref{qab1}), namely the
variables $a_{mn}, b_{mn}$, we obtain the reduced integral for $Q$ in terms
of $a_{mn}^{*},b_{mn}^{*}$ \cite{ple98}. Changing notation, $a_{mn}^{*},
b_{mn}^{*}\to c_{mn},-\bar{c}_{mn}$, the integral becomes:
\bea
&& Q\,\{y\}= \int \prod\limits_{mn}^{} d\bar{c}_{mn}^{}
dc_{mn}^{}\,y_{mn}^{\,2}\,\exp\sum\limits_{mn}^{}\,
\Big[\,y_{mn}^{-2}\,c_{mn}^{}\bar{c}_{mn}^{}\,+
\cr
&& +\,(c_{mn}^{}+\bar{c}_{mn}^{})\,(t_{1}c_{m-1n}^{} -
t_{2}\bar{c}_{mn-1}^{})- y_{mn}^{\,2}\,t_{1}t_{2}\,
c_{m-1n}^{}\bar{c}_{mn-1}^{}\,\Big]\,,
\label{qcc1}
\eea
where $c_{mn},\bar{c}_{mn}$ are again Grassmann variables, and we assume:
$y_{mn}^{\,2}\, \exp\,(\,y_{mn}^{-2}c_{mn}\bar{c}_{mn})= y_{mn}^{\,2}
+c_{mn}\bar{c}_{mn}$, with $y_{mn}^{2}=0,1$.  The integrals (\ref{qab1})
and (\ref{qcc1}) are still the exact expressions for $Q\{y\}$
originally defined in (\ref{qss1}). Taking $y_{mn}=1$ at all sites, we
obtain the 2DIM integral for the pure case:
\bea
&&
Q\,\{1\}= \int \prod\limits_{mn}^{} d\bar{c}_{mn}^{}
dc_{mn}^{}\exp\sum\limits_{mn}^{}\,
\Big[c_{mn}^{}\bar{c}_{mn}^{}+
\cr
&&
+\,(c_{mn}^{}+\bar{c}_{mn}^{})\,(t_{1}c_{m-1n}^{}
-t_{2}\bar{c}_{mn-1}^{})\,
-t_{1}t_{2}\,c_{m-1n}^{}\bar{c}_{mn-1}^{}\,\Big]\,.
\label{qcc1ah}
\eea
In particular, the evaluation of the integral (\ref{qcc1ah}) by
transformation to the momentum space results the Onsager's expression for
$Z$ and $\ln Z$ of the standard rectangular lattice \cite{ple98}.
The advantage of the reduced representation with two variables per site
like (\ref{qcc1}) and (\ref{qcc1ah}) is also that it  explicitly
illuminates the Majorana-Dirac structures of 2DIM already at the lattice
level. This can be most easily seen in the pure case.
\footnote{ \
In the pure case, the lattice Majorana like action for 2DIM
readily follows from (\ref{qcc1ah}) by substitution $c_{m-1n} \to c_{mn}
-\partial_{m}c_{mn},\; \bar{c}_{mn}\to \bar{c}_{mn} -\partial_{n}
\bar{c}_{mn}$, where $\partial_m,\partial_n$ are lattice derivatives
(momenta). This results the Majorana like action $S=\bar{m}
c_{mn}\bar{c}_{mn} +\ldots$ with mass term and the kinetic part
\cite{ple98}.  Evidently, the mass parameter will be $\bar{m} =1-t_1-t_2
-t_1t_2$. The condition $\bar{m}=0$ defines the critical point of 2DIM in
pure case \cite{ple98}. This condition $\bar{m}=0$ may be rewritten as well
in the form: $\sinh(2b_1) \sinh(2b_2) =1$. The disordered phase corresponds
to positive mass, while the ordered phase corresponds to negative mass.}

To prepare the $N$-replicated integral (\ref{qcc1}), with same set of
$y_{mn}$ in each copy, it is suitable to multiply first (\ref{qcc1}) by
factor $\delta(y_{mn} -1) +\delta(y_{mn}-0)=1$, which realizes the
decomposition over the states $y_{mn}=0,1$. We assume that $y_{mn}=1$ is
realized with probability $p$, and $y_{mn}=0$ is realized with probability
$1-p$. The averaging of the $N$-replicated integral (\ref{qcc1}) over the
disorder within the $N$-replica scheme finally results the theory with
interaction presented by the integral \cite{ple98}:
\bea
\Big[Q^N\{y\}\Big]_{\,av}=
\int \prod\limits_{mn}^{}\prod\limits_{\alpha=1}^{N}
d\bar{c}_{mn}^{\;(\alpha)}dc_{mn}^{\,(\alpha)}\,\prod\limits_{mn}^{}
\Big[\,p\,\prod\limits_{\alpha=1}^{N}
\mbox{e}^{\,S_{mn}^{\,(\alpha)}} + (1-p)\,\prod\limits_{\alpha=1}^{N}
c_{mn}^{\,(\alpha)}\bar{c}_{mn}^{\;(\alpha)}\;\Big]\;\;&&
\nonumber\\
=p^{L^2}\!\!\int\prod\limits_{mn}\prod\limits_{\alpha=1}^{N}
d\bar{c}_{mn}^{\;(\alpha)}dc_{mn}^{\,(\alpha)}\,\exp\,
\sum\limits_{mn}^{}\Big[\sum\limits_{\alpha=1}^{N}
S_{mn}^{\,(\alpha)}+ \frac{1-p}{p}\prod\limits_{\alpha=1}^{N}
c_{mn}^{\,(\alpha)}\bar{c}_{mn}^{\;(\alpha)}\mbox{e}^{\,
-S_{mn}^{\,(\alpha)}}\Big]\,, \;\;\;
\label{replica1}
\eea
where $S_{mn}^{\,(\alpha)}$ is the replicated Gaussian action from
(\ref{qcc1}) for the pure case, see (\ref{qcc1ah}). The effect of dilution
is introduced here through the second non-Gaussian term, with bar coupling
constant $g_0 \propto \frac{1-p}{p}$. The continuum-limit field theory for
weak site dilution (RS 2DIM) that follows from the exact lattice integral
(\ref{replica1}) is commented in more detail in \cite{ple98}. This
corresponds to the low-momenta sector of the exact lattice theory
associated with (\ref{qcc1}) and (\ref{replica1}). To extract the effective
low-momenta (long-wave) effective action, one has to distinguish explicitly
the higher and low-momentum lattice fermionic modes in the exact lattice
action (\ref{replica1}). Integrating out the higher-momentum modes in the
first order of perturbation theory then results the $N$-colored Gross-Neveu
model ($N\to0$) with action \cite{ple98}:
\bea
\nonumber
&& S_{\rm\,G-N} =
\int d^2x \Big\{\sum\limits_{\alpha=1}^{N}\Big[\,
m_N\,\psi_{1}^{(\alpha)}\psi_{2}^{(\alpha)}+\,\frac{1}{2}\,
\psi_{1}^{(\alpha)}\,(\partial_1+i\,\partial_2)\,\psi_{1}^{(\alpha)}
\\
\label{gross1}
&& +\, \frac{1}{2}\,\psi_{2}^{(\alpha)}\,(-\partial_1+i\,\partial_2)\,
\psi_{2}^{(\alpha)}\Big] + g_{\mathstrut N} \Big[\sum\limits_{\alpha=1}^{N}
\psi_{1}^{(\alpha)}\psi_{2}^{(\alpha)}\Big]^{\,2}\,\Big\}\,,
\\ \nonumber
&& m_{\mathstrut N}\,
= {1-t_1-t_2-t_1\,t_2 \over \sqrt{2(t_1t_2)_c}}
+ \left<A\right>^N\,\frac{1-p}{p}\,\frac{\left<B\right>}{\left<A\right>}\,
\frac{1}{\sqrt{2\,(t_1t_2)_c}}\,,
\\ \nonumber
&& g_{\mathstrut N}\,=\,\left<A\right>^N\,
\frac{1-p}{p}\,\frac{\left<B\right>^2}{\left<A\right>^2}\,
\frac{1}{4\,(t_1t_2)_c}\,,
\eea
where $\psi_1,\psi_2$ are the anticommuting Majorana components,
$m_N$ and $g_N$ are the effective mass and charge,
respectively. The parameters $\left<A\right>$ and $\left<B\right>$
are some lattice fermionic averages (definite numbers) explicitly
calculated in \cite{ple98}.  Since the replica limit $N \to 0$ is assumed
at final stages, one can put $N=0$ and $\left<A\right>^{N}=1$ in mass and
charge already in (\ref{gross1}). The Gaussian part in (\ref{gross1}) is
the replicated Majorana action, corresponding to the pure case, with the
mass term modified by disorder. The condition of zero mass will give the
coexisting curve $(T_c,p)$ for quenched site dilution, which is exact for
small concentration of vacancies, as $p\to 1$. The analysis of that
curve at strong and moderated dilution, that is coded in the exact
integrals like (\ref{qab1}), (\ref{replica1}), has not yet been performed
in detail. This will probably claim for the advanced methods of
approximation like lattice RG or application of variational approaches
like Hartree-Fock-Bogoliubov.

The effective continuum-limit $N=0$ Gross-Neveu model similar to
(\ref{gross1}), but with another $m_N$ and $g_N$, has been originally
derived and analyzed by DD-SSL as an effective theory near $T_c$ for weak
bond dilution \cite{dots83,shal94,jugshal96, pico06, cckenrul09}. The
DD-SSL predictions for weak bond dilution, based on the renormalization
group (RG) analysis of their effective $N=0$ Gross-Neveu model in the low
momentum sector, with taking also into account some fine symmetry effects
related to the Kramers-Wannier duality and the CFT interpretations of 2DIM
in the pure case, are the double-logarithmic singularity in the specific
heat and the logarithmic corrections to the pure-case power laws in other
thermodynamic functions, as $T\to T_c $ \cite{dots83,shal94,jugshal96,
pico06, cckenrul09}. The derivation of the effective action for site
dilution in the $N=0$ Gross-Neveu form (\ref{gross1}) thus supports the
idea of the double-logarithmic singularity in specific heat and only
logarithmic corrections to the pure-case power laws in other functions (as
in the DD-SSL scheme) also for random-site 2D Ising ferromagnets, for weak
quenched dilution \cite{ple98}. For more details and a recent discussion of
the effects of quenched disorder in RB and RS versions of 2DIM along
theoretical and experimental (Monte-Carlo) lines also see
\cite{shvas01, ssvas98, luo01, pico06, martpla07,hasen08, kenaruiz08,
cckenrul09, fytas08, nihat09, wuzhao09}. In conclusion, we note that the
fermionic integrals like (\ref{qab1}), (\ref{qcc1}) and (\ref{replica1})
are still the exact lattice expressions for $Z$ (either its reduced version
$Q$).  Respectively, one can try other methods of the averaging as well as
other tools of analysis of fermionic theories of random-site 2DIM directly
on a lattice, starting from these exact fermionic integrals.


%
%
\section{The Blume-Capel model }

The Blume-Capel (BC) model is a classical spin-1 model originally
introduced to study phase transitions in specific magnetic materials
with a possible admixture of non-magnetic states. This is a model with
the tricritical point at the critical line in $(T_c,\Delta_{0})$ plane
\cite{beg71,beale86,deng05,lawri84}. The Blume-Capel model can also be
viewed as the annealed site-dilute version of the ordinary Ising model
\cite{cfple08}. The hamiltonian is:
\bea
H = -\sum_{m=1}^{L}\sum_{n=1}^{L}\Big[J_1 S_{mn}S_{m+1n}
+J_2 S_{mn}S_{mn+1}\Big] +\Delta_0\sum_{m=1}^{L}
\sum_{n=1}^{L}S_{mn}^{2}\;, \;\;\;
\label{ham1a}
\eea
where $S_{mn}=0,\pm1$ is the BC spin-1 variable associated with the $mn$
lattice site ($m,n=1,2,3,\ldots,L)$. As distinct from the quenched
disorder case, the zero-spin or vacancy state of $S_{mn}=0,\pm1$ is
now rather to be considered as a one of the three possible states of spin
variable \cite{beg71, beale86, silva06, cfple08}. The hamiltonian modulus
in the Gibbs exponential is:
\bea
-\beta H =\sum_{m=1}^{L}\sum_{n=1}^{L}\Big[K_1 S_{mn}S_{m+1n}
+K_2 S_{mn}S_{mn+1}\Big] +\Delta\sum_{m=1}^{L}
\sum_{n=1}^{L}S_{mn}^{2}\, ,\;\;\;
\label{ham1ab}
\eea
with $K_{1,2} =\beta J_{1,2}$ and $\Delta =-\beta\Delta_0$. The
decomposition like $S_{mn} =y_{mn}\sigma_{mn}$, that was used in the
quenched case, is still possible, but in order to eliminate the Ising
degrees by transformation of $Z$ into a fermionic integral it is
now more suitable to make use rather of the gauge transformation like
$S_{mn} \to \sigma_{mn}S_{mn}$ under the averaging, since the product of
cosines (factor $R\{y\}$) is to be included into $Z$ anyhow, before
averaging. The Boltzmann factors from the Gibbs exponential associated
with (\ref{ham1ab}) can be written as (extended) polynomials in variables
$S_{mn}=0,\pm1$. The polynomial interpretation is important for
fermionization \cite{cfple08}. The partition function becomes (with
$\lambda_{\,i}=\sinh K_{\,i}\,,\, \lambda_{\,i}'=\cosh K_{\,i} -1$):
\bea
Z=\begin{array}[t]{c}
\mbox{Sp}\cr \scriptstyle{(S)}
\end{array}
\Big\{\prod_{m=1}^{L}\prod_{n=1}^{L} e^{\,\Delta S_{mn}^{2}}\,
\Big[
(1+ \lambda_1\, S_{mn}S_{m+1n}
+\lambda_{1}'S_{mn}^{2}S_{m+1n}^{2})
\nonumber \\
\times\,(1+ \lambda_2\, S_{mn}S_{mn+1}
+\lambda_{2}'\, S_{mn}^{2}S_{mn+1}^{2}) \Big]\Big\}.
\label{zet1bcm}
\eea
The factorization of local bond weights can again be performed by analogy
with the Ising case, but now we have to add the even part of the polynomial
into the Gaussian exponential in the measure \cite{cfple08}. For the
horizontal weights, with Grassmann variables $a_{mn},\bar{a}_{mn}$, we
write:
\bea
&&
1+\lambda_1S_{mn}S_{m+1n} +\lambda_1'
S_{mn}^{2}S_{m+1n}^{2}=
\nonumber \\
&&
=\int d\bar{a}_{mn}\,da_{mn} \,
\exp\{(1+\lambda_1'S_{mn}^{2}S_{m+1n}^{2})\,a_{mn}^{}\bar{a}_{mn}\}
\nonumber \\
&&
\times\;(1+a_{mn}S_{mn})\,(1+\lambda_{1}\,\bar{a}_{mn}S_{m+1n})\,,
\label{fact1bcm}
\eea
and similarly we can factorize the vertical bond Boltzmann weights. The
factorization (\ref{fact1bcm}) makes it possible to pass to the purely
fermionic expression for $Z$ in few steps. The final
fermionic integral for $Z$ appears in the form \cite{cfple08}:
\bea
&&
Z=(2e^{\Delta}\cosh K_1 \cosh K_2)^{L^2}
\int \prod\limits_{m=1}^{L}\prod\limits_{n=1}^{L}
d\bar{a}_{mn}da_{mn}d\bar{b}_{mn}db_{mn}
\nonumber \\
&&
\times\; \exp\Big\{\sum\limits_{m=1}^{L}
\sum\limits_{n=1}^{L}
\Big[\,a_{mn}\bar{a}_{mn} +b_{mn}\bar{b}_{mn}
+a_{mn}b_{mn}
\nonumber \\
&&
+(t_1\bar{a}_{m-1n} +t_2\bar{b}_{mn-1})(a_{mn} +b_{mn})
+t_1t_2\,\bar{a}_{m-1n}\bar{b}_{mn-1}
\nonumber \\[1ex]
&&
+\;g_0\;a_{mn}\bar{a}_{mn}b_{mn}\bar{b}_{mn}\,
\exp\,(-\gamma_1a_{m-1n}\bar{a}_{m-1n}
-\gamma_2b_{mn-1}\bar{b}_{mn-1}
\nonumber \\[1ex]
&&
-t_1t_2\,\bar{a}_{m-1n}\bar{b}_{mn-1})\Big]\Big\}\,,
\label{bcap1}
\eea
with parameters (where $\Delta =-\beta\Delta_0$):
\bea
g_0=\frac{e^{-\Delta}}{2 \cosh K_1 \cosh K_2},\;\;\;
\gamma_{i}=1-\frac{1}{\cosh K_{i}}=1-\sqrt{1-t^2_{i}}\,.\;\;\;
\;\;\;\label{bcap2}
\eea
The fermionic integral (\ref{bcap1}) is still the exact expression, even
for finite lattices, provided we assume free boundary conditions both for
spins and fermions. The exponential in the interaction term can be expanded
into a series, which results a finite polynomial in Grassmann variables.
For instance, for particular exponential factor we find: $\exp(-\gamma_1
a_{m-1n} \bar{a}_{m-1n}) =1-\gamma_1 a_{m-1n} \bar{a}_{m-1n}$, and
analogously one can expand other factors. The B-C model is thus presented
in (\ref{bcap1}) as a fermionic theory with free-fermion (Gaussian)
part and a polynomial interaction terms in the action, the highest term in
the interaction polynomial is of order 8 in fermions.  The overall coupling
constant in interaction term is $g_0 \propto \exp\{\beta\Delta_0\}$, the
increasing $g_0 \propto \exp\{\beta\Delta_0\}$ means increasing dilution.
There are also additional parameters $\gamma_{1,2}$ in the interaction
polynomial. These parameters come from accounting properly the weights
related to factors $\cosh(b_1 S_{mn}S_{m+1n}) \cosh(b_2 S_{mn} S_{mn+1})$
in $Z\{S\}$ for Blume-Capel, there are no analogs of $\gamma_{1,2}$
terms in the quenched case. In fact, these terms with $\gamma_{1,2}$
are responsible for the existence of the tricritical point in the
Blume-Capel model. The elimination of the variables like $a_{mn},b_{mn}$ is
now not possible in action (\ref{bcap1}), at least straightforwardly, as
distinct from the quenched case (cf.  (\ref{qcc1})-(\ref{qcc1ah})), just
because of the presence of the combinations like $a_{mn}a_{m-1}$ and
$b_{mn}b_{mn-1}$ in the $\gamma_{1,2}$ terms in the non-Gaussian part of
the lattice action (\ref{bcap1}).

Despite of the non-Gaussian representation for $Z$, it is still possible
to extract physical information by taking the continuous limit (low
momenta sector) of the BC lattice action like (\ref{bcap1}) and analyzing
it using tools from quantum field theory. The details of constructing the
effective two-component fermionic action at low momenta can be seen in
\cite{cfple08}. The resulting action includes the Gaussian part, with mass
term modified by disorder, and the four-fermion interaction of the form
$(\psi\bar{\psi}\,|\,\partial_{x}\psi\partial_{y}\bar{\psi})$.  The
condition of the zero effective mass already gives the
equation for the BC line of phase transitions (critical line) in the
$(T_c,\Delta_0)$ plane, while the effect of the interaction is merely to
modify the kinetic terms in the action \cite{cfple08}. This also provide
grounds to estimate the position of the tricritical point \cite{cfple08}.
These effects are shortly commented below.

The critical line is given by the condition of vanishing the mass term in
effective BC action:  $m_{BC} =1-t_1-t_2 -t_1t_2 +g_0=0$, where $g_0$ is
given in (\ref{bcap2}).  Following \cite{cfple08}, we now consider the
isotropic lattice case, with $t_1=t_2=t$ and $K_1=K_2 =K$. The critical
line is given by $m_{\,BC} =1+g_0 -2t -t^2=0$, with $t=\tanh K$ and
$K=\beta J$, where $\beta=1/T$. In notation with $K =\frac{J}{T} \to
\frac{1}{T}$  and $\frac{\Delta_0}{J} \to \Delta_0$, the criticality
condition becomes:
\bea
\tanh^2\left(\frac{1}{T}\right)+2\tanh\left(\frac{1}{T}\right)
-1=\frac{e^{\frac{\Delta_0}{T}}}{2 \cosh^2 \left(\frac{1}{T}
\right)}\,,\;\;\;
\label{crit1}
\eea
which may be written as well in the form:
\bea
\sinh\left(\frac{2}{T}\right) =1+
\frac{1}{2}\, \exp\left(\frac{\Delta_0}{T}\right)\,, \;\;\;
\label{crit2}
\eea
which in turn admits the explicit solution for $\Delta_0$ as function of
$T=T_c$ in the form:
\bea
\Delta_{0} =T\,\ln\left[2\sinh\left(\frac{2}{T}\right) -2\right]\,.\;\;
\label{crit3}
\eea
The inverse dependence for $T_c$ as function of $\Delta_0$ can be evaluated
numerically by solving any of the above equations, which are all equivalent
to the condition of the zero mass in the effective continuum-limit theory
that follow from (\ref{bcap1}). This results the critical line for the BC
model shown in Fig.~1 in \cite{cfple08}. The critical line is started
with maximal $T_c$ at the left end at $\Delta_0=-\infty$, which corresponds
to the pure case ($g_0=0$), and goes lower as dilution increases, with
increasing $\Delta_0$ and $g_0\propto \exp{\frac{ \Delta_0-2}{T}}$. The
critical line finally terminates at $\Delta_0 =2$ at zero temperature.
There is no ordered phase at stronger dilution,  as it also can be deduced
from (\ref{crit1})-(\ref{crit3}). The theoretical critical line is
compared with the results of the recent Monte-Carlo simulations for B-C
model, see Fig.~1 in \cite{cfple08}. The agreement is found to be very
good (typically within 1\% accuracy) over the whole temperature range
\cite{cfple08}. The available numerical data for $(T_c,\Delta_0)$
are also presented (in part) in Table 1.

The position of the tricritical point at the $(T_c,\Delta_0)$ line can as
well be estimated from the condition of vanishing the kinetic (stiffness)
coefficient in the effective B-C action associated with (\ref{bcap1})
\cite{cfple08}. The Hartree-Fock-Bogoliubov method has been applied to
decouple the four-fermion interaction term in the effective action to
extract the corrections to the kinetic part \cite{cfple08}. The singular
point where the kinetic coefficient vanishes was found at $(T_{t}^{*},
\Delta_{0,t}^{*}) \simeq (0.42158, 1.9926)$, in a reasonably good agreement
with the results of Monte-Carlo simulations for the position of
tricritical point:  $(T_{t}, \Delta_{0,t})\simeq(0.610,1.9655)$
\cite{beale86}, and $(T_t,\Delta_{0,t}) \simeq(0.609(3), 1.966(2))$
\cite{silva06}. It is in general important that the B-C fermionic integral
with a non-Gaussian action (\ref{bcap1}) finally  predicts the existence of
a tricritical point at the B-C critical line at strong dilution, somewhere
close to the termination point of that line at $\Delta_0=2$. The method of
constructing the critical line from the condition of zero mass in fermionic
integral for $Z$ has been recently extended by Fortin and Clusel
\cite{fortclu08} to the general set of the spin-$S$ Ising models ($S=1/2,
1, 3/2, 2, 5/2, \ldots\,$) \cite{fortclu08}. The standard 2DIM and the
Blume-Capel models are the first two representatives in this set. The
agreement of the theoretical predictions for the critical line with the
available Monte-Carlo data for the spin-$S$ models was again found to be
very good, even despite of highly complicated polynomial structures, with
many fermions, arising  for the higher spin-$S$ Ising models in the kinetic
part of the action \cite{fortclu08}. These features may be probably
understood as an evidence for the well expressed clusterization processes
at low temperatures in such models, including the generic case of the
spin-1 Blume-Capel model.

 
%
%

\begin{table}[!tb]
\centering
\begin{tabular}{ c | c c c }
\hline
Chemical potential & & Temperature $T_c(\Delta_0)$ &
\\
$\Delta_0$
& Ref. \cite{beale86} & Ref. \cite{silva06}
& Eqs. (\ref{crit1})--(\ref{crit3})
\\
\hline
\\
0.0   & 1.695 & 1.714(2) & 1.6740
\\
0.5 & 1.567 & 1.584(1) & 1.5427
\\
1.0 & 1.398 & 1.413(1) & 1.3695
\\
1.5 & 1.150 & 1.155(1) & 1.1162
\\
1.87 & 0.800 & 0.800(3) & 0.7712
\\
1.95  & 0.650 & 0.651(2) & 0.6135
\\
1.962 & 0.620 & 0.619(1) & 0.5776
\\
1.969 & 0.600 & 0.596(5) & 0.5531
\\
\hline
\end{tabular}
\caption{
Numerical values of the critical points $(T_c(\Delta_0),\Delta_0)$ in the
Blume-Capel model: comparison of the results of Monte-Carlo simulations
and the equations (\ref{crit1})--(\ref{crit3}). Note that small variation
of $\Delta_0$ causes more significant changes in $T_{c}(\Delta_0)$ in the
region near $\Delta_0=2$, as it is to be expected from
(\ref{crit1})--(\ref{crit3}).}
\end{table}

%
%
\section{Conclusions}

The integrals with anticommuting (Grassmann) variables provide effective
tools to analyze pure and disordered Ising like spin models in two
dimensions. The Ising spin glasses, geometry disordered lattices, regularly
diluted models, also can be analyzed along these lines. In a more general
context, it may be noted that there are as well few other important
physical problems with spins and fermions in two dimensions. The most
prominent are the quantum Hall effect and the high-$T_c$ superconductivity
in oxide cuprates and related substances. It is interesting that the fine
tuning in spin-fermion correspondence as well as the effects of disorder
seemingly play the role in both cases. It is therefore important to
understand better the ordering phenomena in terms of fermions
in such systems.


%
%

%
%

\end{document}